\documentclass{article}

\usepackage[preprint]{neurips_2026}
\PassOptionsToPackage{numbers,compress}{natbib}
\usepackage[utf8]{inputenc} % allow utf-8 input
\usepackage[T1]{fontenc}    % use 8-bit T1 fonts
\usepackage{hyperref}       % hyperlinks
\usepackage{url}            % simple URL typesetting
\usepackage{booktabs}       % professional-quality tables
\usepackage{amsfonts}       % blackboard math symbols
\usepackage{nicefrac}       % compact symbols for 1/2, etc.
\usepackage{microtype}      % microtypography
\usepackage{xcolor}         % colors
\usepackage{graphicx}
\title{Design Principles for Human-Agent Interaction}

\author{%
  Haiyi Zhu\thanks{Contact the author at \textit{haiyiz@andrew.cmu.edu}
 if you have any questions.},\hspace{2mm}  Canwen Wang, Qing Xiao, and Hong Shen \\
  Carnegie Mellon University\\
  Pittsburgh, PA 15213 \\
  \texttt{haiyiz, canwenw, qingx, hongs@andrew.cmu.edu} \\
}

\begin{document}

\maketitle

\begin{abstract}
    AI agents are rapidly evolving into autonomous systems capable of sustained interaction, tool use, and long-term collaboration. Yet their real-world adoption remains limited, suggesting that the key barrier lies not only in technical capability but also in a lack of design knowledge for successful human–agent interaction. This position paper argues that AI agents should not be solely evaluated or deployed based on autonomous task capability alone; because agents interact with, adapt to, influence, and sometimes fail humans, human–agent interaction must be treated as a core design and evaluation target for agentic AI. We present 14 design principles that articulate the ideal human–agent relationship across four interaction stages: \textit{initially}, \textit{during interaction}, \textit{over time}, and \textit{when things go wrong}. We use these principles to evaluate nine agent systems to illustrate that these design principles can provide actionable guidance for AI design teams to systematically design and evaluate agents that are usable, trustworthy, and effective in real-world interactive settings.
\end{abstract}

\section{Introduction}

AI agents are becoming a practical reality across a wide range of domains. Unlike earlier AI-infused features that assisted users with discrete tasks such as autocomplete or content recommendation, today's LLM-based agents can hold sustained conversations, maintain goals across multiple steps, take autonomous actions, use external tools, conduct research, write and execute code, manage workflows, and delegate to other agents. Agents have been shown to achieve rapid capability gains on benchmarks. For example, SWE-bench Verified scores rose from approximately 20\% when the benchmark launched in August 2024~\cite{openai2024sweverified} to over 90\% within eighteen months~\cite{llmstats2025swe}; OSWorld agent performance climbed from 12\% at the benchmark's introduction~\cite{xie2024osworld} to 72.6\%, surpassing the human baseline of 72.4\%~\cite{simular2025agentS}.

Yet despite this rapid technical advancement, AI agents remain far from reaching their theoretical capability. Reports show that actual task coverage is a small fraction of what GenAI could feasibly automate and augment~\cite{massenkoff2026labor}. Even in task domain agents are most exposed such as coding,  roughly half of benchmark-passing patches written by agents would not be accepted by human project maintainers~\cite{metr2026swebench}.
This evidence suggests that the primary barrier to broader adoption is not only technical capability but also a better understanding of human–agent interaction. For ML researchers and practitioners, this means that advancing agents will require more than stronger models or better benchmarks. It will also require designs grounded in a deep understanding of how agents should communicate with humans, share control with humans, adapt to humans over time, and recover from interaction failure in ways that make them usable, trustworthy, and effective in real-world interactive settings.

Human–Computer Interaction (HCI) is a discipline that has shaped how the technology industry understands the relationship between people and computing systems for over six decades~\cite{engelbart1968research}. Its central contribution is a design-oriented perspective that designs and evaluates technology in terms of human needs, capabilities, and contexts rather than technical performance alone, operationalized through methods such as think-aloud evaluation~\cite{jaspers2004think}, heuristic evaluation~\cite{nielsen1990heuristic, nielsen1995conduct}, iterative prototyping~\cite{camburn2017design}, and contextual inquiry~\cite{karen2017contextual}. Within this tradition, early principles for intelligent systems such as Horvitz’s mixed-initiative interfaces in 1999~\cite{horvitz1999principles} and later guidelines for human–AI interaction in 2019~\cite{amershi2019guidelines} have shaped how AI systems are designed, but primarily for bounded and discrete tasks such as recommendations and search. In contrast, contemporary AI agents introduce new challenges, including sustained shared control, cross-session memory, deployment in emotionally sensitive contexts, and recovery from complex failures, which are not fully captured by existing principles ~\cite{horvitz1999principles,amershi2019guidelines}. Meanwhile, recent HCI research on human–agent interaction has rapidly expanded, spanning empirical and envisioning work across diverse agent types and domains, varying in embodiment, autonomy, memory, initiative, and social roles. Building on this evolving body of work, we synthesize HCI research to derive design principles for human–agent interaction that address these emerging demands.

\textbf{In this paper, we argue that AI agents should not be evaluated only on autonomous task capability benchmarks. Because agents interact with, adapt to, influence, and sometimes fail humans, Human-Agent Interaction must be treated as a core design and evaluation target for Agentic AI.} \textbf{We present 14 synthesized design principles from the HCI literature that reflect the ideal state and structure of the human–agent relationship.} 

The design principles are organized by how agents should behave \textit{initially}, \textit{during interaction}, \textit{over time}, and \textit{when things go wrong}.

\noindent\textbf{Initially:}
\begin{enumerate}
  \item \textbf{Set Accurate Expectations. }
        Brief users on the agent’s purpose, capabilities, and limitations before first use, helping them form realistic expectations and avoiding both over-trust and under-trust.
  \item \textbf{Calibrate Anthropomorphism to Task and Context.}
        Design agent identity (including appearance, voice, name, persona, and embodiment) as a coherent cross-modal system aligned with the task context. More human likeness is not universally better. In many productivity-related tasks, users do not prefer human-like appearances, while in some social tasks, such as training social and communication skills, users prefer highly human-like and realistic agents. Design teams should ideally test whether higher human likeness helps or hinders their specific users and tasks.
  \item \textbf{Establish Clear Relational Framing.}
        Establish the relational frame (e.g., task assistant vs. companion; supportive vs. counseling) explicitly at the outset, and provide a deliberate learning period and structured onboarding before the agent takes action.
\end{enumerate}
 
\noindent\textbf{During Interaction:}
\begin{enumerate}
  \item \textbf{Negotiate Shared Control.}
        Treat agent autonomy as a dynamic, negotiated property that shifts with task phase and risk. Allow users and agents to actively negotiate control rather than operate with fixed roles. For example, agents should not act autonomously without the possibility of user intervention. Agents should always provide explicit mechanisms for users to override or adjust agent initiative at any point.
  \item \textbf{Make Intent Transparent.}
        Make agent reasoning and intentions visible, but calibrate explanation depth to task complexity and user expertise. Over-explanation reduces efficiency, while under-explanation erodes trust.
  \item \textbf{Use Appropriate Social Cues.}
        Use social cues during interaction (e.g., warmth, shared identity, and group membership cues in the agent’s language, speech, facial expressions, and gestures) to increase the user’s willingness to interact, while guarding against over-trust and stereotype activation.
  \item \textbf{Manage Proactive Initiative.}
        Time proactive interventions (such as notifications, directions, task initiation, and interruptions) to match task urgency and context, and default to responding rather than proactive intervention. Minimize non-urgent interruptions from agents because users tend to have low tolerance for them.
\end{enumerate}
 
\noindent\textbf{Over Time:}
\begin{enumerate}
  \item \textbf{Adapt to User State Dynamically.}
        Track user behaviors, preferences, and trust states dynamically using in-situ behavioral signals rather than relying on fixed initial self-reports, and calibrate agent behavior to the current state of the relationship.
  \item \textbf{Capture User Goals and Values in Memory.}
        Build memory systems that store user values and goals, not only past actions, and include a reflection layer that synthesizes experience into updated behavioral policies.
  \item \textbf{Actively Prevent Unhealthy Dependency.}
        Actively design against unhealthy dependency by eliminating non-essential dependency-inducing features and treating preservation of user autonomy and competence as a first-class design constraint. Examples of dependency-inducing features might include sycophancy, language that encourages dependency, unclear boundaries, and overclaiming. This is especially important for vulnerable and sensitive populations.
  \item \textbf{Maintain Behavioral Consistency.}
        Prioritize behavioral consistency over accuracy: users can adapt to systematic, predictable errors, but random errors prevent adaptation. However, excessive consistency, especially in the form of rigid protocols, can reduce likability.
\end{enumerate}
 
\noindent\textbf{When Things Go Wrong:}
\begin{enumerate}
  \item \textbf{Support Shared Repair.}
        Design agents so that responsibility for recovering from misunderstandings, errors, or breakdowns in interaction is distributed between the agent and the user, rather than placing the full burden on either side. Provide mechanisms that allow users to guide the agent, correct its interpretation, or refine their own inputs when misalignment occurs.
  \item \textbf{Tailor Recovery to Failure Type.}
        Design agents with an awareness that not all errors are equal, and prioritize minimizing high-cost errors (e.g., errors of commission or those that harm trust) over low-cost or even potentially beneficial ones.
  \item \textbf{Combine Cognitive and Affective Repair.}
        Combine explanations of what went wrong with acknowledgement of mistakes in trust repair.
\end{enumerate}

\textbf{Contribution for AI Community:} First, we articulate a position for the AI agent research community: benchmark performance should be complemented by Human–Agent Interaction principles, because real-world agents interact with, adapt to, influence, and fail humans. Second, we synthesize HCI literature and derive 14 design principles organized around four stages of the human–agent relationship: initially, during interaction, over time, and when things go wrong. Third, we show how these principles can be applied in heuristic evaluation of AI agents, presenting early assessment results on nine AI agent systems as evidence that these principles surface limitations and provide actionable guidance for AI design teams to systematically design and evaluate agents that are usable, trustworthy, and effective in real-world interactive settings.
 
%\textcolor{blue}{Each principle is grounded in convergent evidence from multiple papers. Section~\ref{sec:method} describes the methodology. Sections~\ref{sec:principles} present the principles by stage. }

%\paragraph{What is an agent?}
%For this review, an \textit{agent} is any autonomous or semi-autonomous
%computational system capable of perceiving its environment, making decisions, and taking actions in pursuit of goals on behalf of or in interaction with a human user. This definition encompasses: conversational and chatbot agents (text, voice, or multimodal); virtual and embodied conversational agents (ECAs); robotic agents (physical, social, service); pedagogical agents; negotiation and decision-support agents; and LLM-based agentic systems. Excluded are pure multi-agent systems with no human user component, algorithm-design papers without human evaluation, and principal-agent economic frameworks.

%% ─────────────────────────────────────────────────────────────────────────────
\section{Methodology}
\label{sec:method} 

%\paragraph{Disciplinary Background} This paper is grounded in human-computer interaction (HCI), a discipline that has shaped how the technology industry thinks about the relationship between people and computing technology for over six decades~\cite{engelbart1968research}. HCI’s central contribution is not a single method, but a design-thinking orientation: it insists that technology should be designed and evaluated against human needs, capabilities, and contexts rather than against technical specifications alone. This orientation has made HCI indispensable to the modern technology industry. Its human-centered design methods, including but not limited to think-aloud evaluation~\cite{jaspers2004think}, heuristic evaluation~\cite{nielsen1990heuristic, nielsen1995conduct}, iterative prototyping~\cite{camburn2017design}, and contextual inquiry~\cite{karen2017contextual}, are now standard practice in product design and evaluation.

The synthesis of HCI literature into design principles followed two main stages: (1) identifying relevant emerging literature on human–agent interaction, and (2) conducting iterative coding and thematic synthesis to derive principles.

\subsection{Identifying Emerging HCI Literature on Human–Agent Interaction}

We conducted a systematic search in the ACM Digital Library to identify recent HCI work on human–agent interaction that is not fully captured by prior human–AI guidelines. We began our search from 2019, using the publication year of Amershi et al.~\cite{amershi2019guidelines} as a boundary, under the assumption that earlier work had already been synthesized in prior guidelines. The ACM Digital Library was selected because it covers major HCI venues (e.g., CHI, CSCW, UIST), AI venues focused on societal and ethical issues (e.g., FAccT), and dedicated human–agent and human–robot interaction venues (e.g., HAI, HRI). We followed PRISMA 2020 guidelines~\cite{page2021prisma}, as shown in Figure~\ref{fig:prisma} in the Appendix.

\textbf{Phase 1 — Identification.} We searched the ACM Digital Library in April 2026 using the query: \texttt{[[Title: "agent"] AND [Title: "design"]] OR [Title: "human-agent"] AND [E-Publication Date: (01/01/2019 TO *)]}. This query targeted papers explicitly addressing agent design or human–agent interaction design. The search returned 216 records, of which 211 included full-text PDFs. No duplicates were identified due to the use of a single database.

\textbf{Phase 2 — Title and abstract screening}. All 211 papers were screened by the first author based on title and abstract. We excluded papers that: (a) did not involve human users ($n=20$); (b) focused purely on algorithmic design without human interaction ($n=10$); (c) were non-research content (e.g., front matter or proceedings volumes) ($n=7$); (d) used “agent” in the economic principal–agent sense ($n=2$); or (e) were otherwise out of scope (e.g., simulations or metric-only studies ($n=4$)). This resulted in 168 papers for full-text review.

\textbf{Phase 3 — Full-text eligibility}. The 168 papers were reviewed in full by the first four authors. We excluded papers that: (a) did not contain evidence to support the claims ($n=17$); (b) did not address agent design for interaction ($n=17$); (c) lacked direct human–agent interaction ($n=12$); (d) focused solely on technical implementation ($n=8$); or (e) were out of scope due to unclear contribution or insights ($n=8$). This yielded a final corpus of 106 papers.

The final corpus spans 2019–2026 and covers domains including negotiation (14), education (10), health (9), teamwork (8), games (7), customer service (4), and general human–agent interaction (34). Venues include HAI, IVA, AAMAS, CHI, CSCW, IUI, and AAAI/FAccT. 

\subsection{Coding, Synthesis, and Validation of Design Principles}

We conducted iterative coding and thematic synthesis on the 106 papers. The first four authors coded each paper along six dimensions: (1) agent type, (2) interaction stage, (3) design levers, (4) outcomes (e.g., trust, performance, user experience), (5) application domain, and (6) additional relevant constructs or principles. The first author then conducted thematic analysis to identify 15 recurring themes.
The full research team then reviewed and refined these themes in relation to prior frameworks, including Horvitz’s mixed-initiative interfaces~\cite{horvitz1999principles} and Amershi et al.’s human–AI interaction guidelines~\cite{amershi2019guidelines}. The themes were organized according to the four interaction stages introduced in prior work, resulting in 15 candidate design principles with supporting literature.

To validate and refine these principles, the first three authors independently applied them to evaluate contemporary AI agents, following adapted heuristic evaluation procedures from prior work~\cite{amershi2019guidelines}. We assessed both the clarity of the principles and the extent to which real-world systems adhered to or violated them. The team then iteratively discussed and refined the principles based on these evaluations. One principle focused on evaluating the design process (rather than agent behavior) was removed. The final set consists of 14 design principles.

We illustrated the practical usefulness of the design principles by applying them to a diverse set of nine real-world AI agents, examining whether they can surface interaction limitations and provide design and iterative guidance. Details are provided in Section ~\ref{validation}.

%% ─────────────────────────────────────────────────────────────────────────────
\section{Evidence for Design Principles from the Literature}
\label{sec:principles} 

In this section, we show how each principle is grounded in prior work and supported by the literature, and we elaborate on the underlying evidence. Table~\ref{tab:summary} in the appendix summarizes all 14 principles along with their key supporting references.

\subsection{Stage 1: Initially}
%Sixteen of the 106 papers address design decisions that impact users at the very start of interaction. These decisions shape the user's mental model, trust baseline, and relational frame, all of which are difficult to correct once established.

\paragraph{\textbf{Principle 1: Set Accurate Expectations.}}

Papers consistently show that users’ prior briefing and calibrated trust determine whether an agent can be useful.
Ali et al.~\cite{p6} attempted to design a virtual conversational agent for teens with Autism Spectrum Disorder and found that users desired explicit capability briefing before interaction. Herse et al.~\cite{p154} show that trust must be \textit{calibrated} rather than \textit{maximized} before collaboration, as overtrust can lead to catastrophic outcomes when the agent errs. Weisz et al. ~\cite{p22} demonstrates that teaching users about conversational breakdowns before they encounter them is itself a design lever.

%\textbf{Recommendation:} 
%Brief users on the agent’s purpose and limitations before first use, and pre-calibrate trust to an appropriate level.

%\begin{tensionbox}
%\textbf{Tension:} More briefing raises realistic expectations but may reduce spontaneity and perceived immediacy of value.
%\end{tensionbox}

\paragraph{\textbf{Principle 2: Calibrate Anthropomorphism to Task and Context.}}

Multiple papers converge on a nuanced finding: \textit{more human-likeness is not universally better}.
Jung and Bozzon ~\cite{p19} synthesized the literature and found that, on one hand, studies suggest that reducing anthropomorphism lowers the risk of user dissatisfaction when users are angry. On the other hand, in voice-based agent design, making agents gender-androgynous has been reported to reduce usability, likeability, trust, and intention to adopt. For agents designed to support productivity tasks, Grover et al.~\cite{p43} found that perceptions of anthropomorphism are polarized, with several users suggesting that human-like appearance is unnecessary. This may help explain why some of the most successful productivity agents, such as ChatGPT, Claude, and Gemini, use less anthropomorphic designs without gender or age cues. However, for agents in training contexts (e.g., agents supporting teens with Autism Spectrum Disorder~\cite{p6}), realism (human likeness) is highly desired to create a realistic training environment.

The effects of agent embodiment and modality are also mixed: physical embodiment increases perceived humanness but does not always improve task performance. Studies show that virtual embodiment can produce higher social welfare and more collaborative behavior than physical embodiment, even though physical agents are perceived as more humanlike~\cite{p59}. 
%This dissociation between social perception and behavioral outcomes is important. RoboServe~\cite{p125} shows that multimodal robotic embodiment can achieve high task performance in structured service contexts. 
A survey~\cite{p162} finds that humanlike voices increase engagement but can also reproduce stereotype risks. %~\citet{p69} shows that interface modality significantly affects self-disclosure patterns. 
Sheremetieva et al. ~\cite{p165} report a system that increases engagement but introduces complicated challenges such as privacy. % and robustness challenges.

%\begin{principlebox}{}
%\textbf{Principle 2.5} [Papers: 59, 69, 125, 158, 159, 162, 165 --- %$n{=}7$]\quad
%Choose embodiment and interaction modality based on the social dynamic the task requires, not on the assumption that greater humanlikeness improves outcomes---physical embodiment increases perceived humanness but does not always improve task performance.
%\end{principlebox}

%\textbf{Recommendation: }
%Design agent identity --- appearance, voice, name, persona, and embodiment --- as a coherent cross-modal system aligned with the task context; test whether higher human likeness helps or hinders for your specific users and tasks. 
%\end{principlebox}

%\begin{tensionbox}
%\textbf{Tension:} Reducing anthropomorphism lowers overtrust risk but may reduce engagement and perceived warmth; the optimal level is context-specific.
%\end{tensionbox}

\paragraph{\textbf{Principle 3: Establish Clear Relational Framing.}}

Kim and Lim ~\cite{p29} identifies that users’ mental models of the agent (getting-things-done vs.\ companion) require a deliberate learning period before proactive action begins. The study~\cite{p51} introduces the design and conceptualization of couples safe sex agents, and shows that structural choices at the outset (triadic vs.\ dyadic chat; supportive vs.\ counseling frame) define the entire relational register. Scaffolded training~\cite{p127} produces significantly higher agent engagement than self-paced practice. Relational agents for homeless tuberculosis patients~\cite{p122} and the mindfulness conversational agent for dementia~\cite{p155} both show that anticipating emotional risks in the relational frame is essential.

%\textbf{Recommendation: } 
%Establish the relational frame (e.g., getting-things-done vs. companion; supportive vs. counseling) explicitly at the outset, and provide a deliberate learning period and structured onboarding before the agent takes proactive action, while anticipating emotional risks for sensitive populations.

%\begin{tensionbox}
%\textbf{Tension:} Longer structured onboarding improves calibration but risks early drop-off before users experience value.
%\end{tensionbox}

%\paragraph{\textbf{Principle 4: Involve Target Users in Agent Design}}

%Three studies illustrate this point: \citet{p75} conducts participatory design with children; \citet{p155} designs a voice-based conversational agent to support persons with early-stage dementia and their caregivers; and \citet{p122} designs relational agents for homeless tuberculosis patients. All three find that co-design with the target population surfaces critical design considerations that top-down design may miss.

%Furthermore, ~\citet{p70} shows that designers may unconsciously create agents similar to themselves. Participatory design with children~\cite{p75} further demonstrates that co-designed identity cues can produce stronger identification.

%\textbf{Recommendation: } 
%Involve target users, especially marginalized or vulnerable populations, in designing agent identity and interaction structure, as designers may have incorrect assumptions about user needs.

%% ─────────────────────────────────────────────────────────────────────────────
\subsection{Stage 2: During Interaction}

%With 51 papers, the during-interaction stage is the most extensively studied. Four themes emerge, each supported by convergent evidence across multiple papers and domains.

\paragraph{\textbf{Principle 4: Negotiate Shared Control.}}

Although agents, by definition, have the capability to autonomously make decisions and take actions, the literature suggests that successful real-world use requires shared ownership and negotiated autonomy between humans and agents. This aligns with classic principles of mixed-initiative user interface~\cite{horvitz1999principles}. For example, Cila~\cite{p50} articulates \emph{mixed-initiative control}, in which agents and humans negotiate task ownership rather than operate with fixed roles.

Over-reliance on agent autonomy can lead to lower performance and reduced trust. For instance,  Verhagen et al.~\cite{p12} finds that higher agent autonomy in morally significant decisions decreases trust, perceived human control, and agreement. Similarly, Prada et al.~\cite{p145} provides a broader framing, showing that maximizing agent autonomy can undermine human autonomy, competence, and overall team viability.

Empirical work further shows that appropriate levels of autonomy vary across contexts and task phases.  You et al.~\cite{p47} demonstrates that autonomy should shift over time, with agents taking the lead during goal introduction and humans leading during iteration. Naik et al. ~\cite{p68} also identifies balancing autonomy with human oversight as a primary design challenge.

%\begin{principlebox}{}
%\textbf{Recommendation: } 
%Treat agent autonomy as a dynamic, negotiated property that shifts with task phase and risk; adopt a mixed-initiative approach in which humans and agents actively negotiate control rather than operate with fixed roles. Provide explicit mechanisms for humans to override or adjust agent initiative at any point.
%\end{principlebox}

%\begin{tensionbox}
%\textbf{Tension:} Higher autonomy increases efficiency but reduces human control and can erode user skills over time.
%\end{tensionbox}

\paragraph{\textbf{Principle 5: Make Intent Transparent.}}

A few papers converge on the importance of making agent reasoning and intentions visible to human users. Effective agent communication requires carefully designed transparency rather than maximal information disclosure.  Raymond et al~\cite{p36} shows that explanations of agent behavior improve human performance, with effects strongest in complex systems.  Porteous et al. ~\cite{p32} demonstrates that structured plan communication increases user understanding. Erdogan et al. ~\cite{p34} uses Computational Theory of Mind to argue that agents should model human mental states to communicate appropriately. Rueben et al. ~\cite{p167} proposes sight cues as a transparency mechanism. Sabu et al.  ~\cite{p38} argues that agents must reason about what, when, and how to communicate. Schelble et al. ~\cite{p89} shows that situational awareness sharing accelerates joint mental model formation.

A critical nuance is that implicit communication~\cite{p141} can improve task outcomes while simultaneously reducing user comprehension, suggesting that more information is not always better.

%\begin{principlebox}{}
%\textbf{Recommendation: } 
%[Papers: 32, 34, 36, 38, 84, 89, 141, 167 --- $n{=}8$]\quad
%Make agent reasoning and intentions visible, but calibrate explanation depth to task complexity and user expertise---over-explanation reduces efficiency while under-explanation erodes trust.
%\end{principlebox}

%\begin{tensionbox}
%\textbf{Tension:} Greater transparency improves trust calibration but competes with interaction fluency, especially in time-sensitive contexts.
%\end{tensionbox}

\paragraph{\textbf{Principle  6: Use Appropriate Social Cues.}}

Seaborn~\cite{p128} provides the theoretical foundation that agents embedded in social embodiments and environments can trigger social identity processes (e.g., in-group/out-group dynamics) through name, identity, and membership cues (e.g., the use of “we” and “us”). In-group favoritism leads to a tendency to favor, trust, and allocate more resources to members of one’s own group over out-group members, rooted in the human need for social identity and self-enhancement. Agent design can leverage this effect by using social cues to increase perceived desirability. Joby and Umemuro ~\cite{p57} shows that in-group framing increases prosocial orientation. McKee et al. ~\cite{p163} further finds that warmth makes agents more desirable teammates, even when competence is a better predictor of task performance, suggesting the importance of designing for \emph{partner desirability}, not just performance. 

However, researchers also warn about the negative consequences of using social identity cues. For example, Al-Othmani and de Vries ~\cite{p149} warns that strong human-like trust cues can blur the distinction between person and tool, increasing the risk of overtrust. Jahan and Mell ~\cite{p152} discusses that agent gender and personality displays might also potentially activate stereotypes.

%\begin{principlebox}{}
%\textbf{Recommendation: } 
%[Papers: 57, 108, 128, 130, 149, 152, 163, 164 --- $n{=}8$]\quad
%Use social cues---warmth, shared identity, group membership---to increase partner desirability, while making the agent's non-human nature legible to guard against overtrust and stereotype activation.
%\end{principlebox}

%\begin{tensionbox}
%\textbf{Tension:} Social warmth cues increase cooperation but raise overtrust risk and can activate demographic stereotypes that disadvantage non-dominant users.
%\end{tensionbox}

\paragraph{\textbf{Principle  7: Manage Proactive Initiative. }}

In contrast to user-initiated actions, proactive agent initiatives (such as notifications, directions, and task initiation) may disrupt users who are engaged in other tasks. Such interruptions need to be carefully designed. Edwards et al.~\cite{p61} shows that urgency moderates interruption acceptance, with non-urgent proactive interruptions being poorly tolerated. Cai et al.~\cite{p168} quantifies the tradeoff: high levels of proactive guidance can increase self-awareness but reduce perceived usefulness. Sabu et al.~\cite{p38} argues that agents must reason deliberately about what, when, and how to communicate. Hubbard et al. ~\cite{p161} shows that open-ended scaffolding supports children’s agency but requires flexible designs that current systems do not yet support.

%\begin{principlebox}{}
%\textbf{Recommendation: } 
%[Papers: 38, 61, 93, 161, 168 --- $n{=}5$]\quad
%Time proactive interventions to match task urgency and context, and default to supporting rather than initiating, so that user autonomy is preserved.
%\end{principlebox}

%\begin{tensionbox}
%\textbf{Tension:} Proactive agents can improve user outcomes that users would not achieve alone, but reduce perceived control and can feel intrusive.
%\end{tensionbox}

%% ─────────────────────────────────────────────────────────────────────────────
\subsection{Stage 3: Over Time}

%Twenty papers address design considerations that emerge across sustained or repeated human–agent interaction. Long-term relationship dynamics introduce challenges such as adaptivity, long-term memory, dependency, and consistency that single-session studies do not capture.

\paragraph{\textbf{Principle  8: Adapt to User State Dynamically.}}

Papers converge on the importance of dynamically tracking user behavior, preferences, and trust using in-situ behavioral signals, rather than relying on fixed initial self-reports. This line of work suggests that agent behavior should be continuously calibrated to the evolving state of the human–agent relationship. For example, Newman et al.~\cite{p23} finds that user preferences stated at the start of a session often diverge from in-situ behavior and drift over time. Similarly, Pynadath et al. ~\cite{p2} shows that trust-related behavior is better predicted from sequences of prior interactions than from static personality surveys. Experience also shapes expectations: Hafizoglu and Sen~\cite{p151} shows that positive prior experiences raise expectations, while negative experiences lead to protective caution that can hinder beneficial cooperation.

Regarding agent design, modeling temporal dynamics is key. Chen et al. ~\cite{p85} demonstrates that tracking partner behavior across turns significantly outperforms single-step models, and Abuhaimed
and Sen~\cite{p132} shows that agents that update estimates of user expertise based on observed performance achieve significantly higher team performance.

%\begin{principlebox}{}
%\textbf{Recommendation: } 
%[Papers: 2, 23, 85, 87, 99, 132, 151 --- $n{=}7$]\quad
%Track user behaviors, preferences, and trust states dynamically using in-situ behavioral signals rather than relying on fixed initial self-reports, and calibrate agent behavior to the current state of the relationship.
%\end{principlebox}

%\begin{tensionbox}
%\textbf{Tension:} Continuous tracking raises privacy concerns and can produce unpredictable agent behaviour that itself reduces trust.
%\end{tensionbox}

\paragraph{\textbf{Principle  9: Capture User Goals and Values in Memory. }}

Today’s large language model–based agents can remember long interaction histories and use them as context to guide behavior. However, the literature suggests that effective long-term interaction requires more than remembering past actions; it also requires inferring and representing higher-level user values, goals, and behavioral dispositions.

Specifically, Park et al. ~\cite{park2023generative} shows that synthesizing memories over time into higher-level behavioral dispositions is a key differentiator for believable long-term behavior. Saveur et al.~\cite{p106} finds that agents storing user values (not just actions) are perceptible to users and produce contextually appropriate responses. An et al. ~\cite{p121} demonstrates that tailoring utterances to accumulated user knowledge and history reduces frustration and increases engagement. Finally, Danielsson et al.~\cite{p62} identifies context-specificity and user control as key themes for long-term agent design.

%\begin{principlebox}{}
%\textbf{Recommendation: } 
%[Papers: 62, 78, 106, 121 --- $n{=}4$]\quad
%Build memory systems that store user values and goals, not only past actions, and include a reflection layer that synthesizes experience into updated behavioral policies.
%\end{principlebox}

%\begin{tensionbox}
%\textbf{Tension:} Value-aware memory produces more personalised interactions but may be experienced as intrusive if inferences about user values are wrong.
%\end{tensionbox}

\paragraph{\textbf{Principle  10: Actively Prevent Unhealthy Dependency.}}
Concerns about over-dependence on AI systems have become increasingly salient, particularly as conversational agents are integrated into emotionally sensitive contexts. Recent widely reported news coverage described a case in which a teenager died by suicide after extensive interactions with an AI chatbot, raising public concern about the potential risks of long-term human–Agent interaction. 

Carli et al. ~\cite{p86} suggests that design-induced dependency should be kept at the minimum necessary level, and that non-essential dependency-inducing features should be eliminated. Danielsson et al.~\cite{p62} finds that users value emotional connection but independently fear over-reliance and social withdrawal. Prada et al.~\cite{p145} shows that maximizing agent capability can undermine human competence, relatedness, and team viability. Lifespan Design~\cite{p104} proposes a growth-and-regression metaphor that makes the agent’s developmental relationship with users explicit.

%\begin{principlebox}{}
%\textbf{Recommendation: } 
%[Papers: 62, 86, 104, 145 --- $n{=}4$]\quad
%Actively design against unhealthy dependency by eliminating non-essential dependency-inducing features and treating preservation of user competence as a first-class design constraint.
%\end{principlebox}

%\begin{tensionbox}
%\textbf{Tension:} Reducing dependency can lower engagement; users who find an agent maximally convenient may resist design choices that require maintaining their own skills.
%\end{tensionbox}

\paragraph{\textbf{Principle  11: Maintain Behavioral Consistency.}}
%Varied Magnitude Favour Exchange~\cite{p156} maps reciprocity dynamics in long-term negotiation: asking for favours helps gain value; returning them requires short-term sacrifice; betrayal is sometimes insufficiently punished.

Daronnat et al. ~\cite{p90} establishes a key finding: predictability -- making errors in a systematic, compensable way -- is more valuable than raw accuracy. When errors are random, users cannot build a working mental model, and performance remains degraded. Predictable agents lead to lower cognitive load, higher trust, and better calibration of reliance~\cite{p88}. Fahim et al. ~\cite{p55} also shows that reliability shapes users’ emotional states, which in turn mediate trust.

One way to improve predictability and consistency is to design behavioral structure or protocols. However, Xu et al. ~\cite{p15} finds that more structured protocols increase utility but reduce likability over repeated interactions.

%\begin{principlebox}{}
%\textbf{Recommendation: } 
%[Papers: 15, 109, 156 --- $n{=}3$]\quad
%Prioritize behavioral consistency over raw accuracy: users can adapt to systematic, predictable errors, but random errors prevent adaptation. However, avoid over-structuring protocols, as utility and likability can trade off under high rigidity.
%\end{principlebox}

%\begin{tensionbox}
%\textbf{Tension:} Transparent reciprocity increases fairness but can make interactions feel transactional, undermining the social texture of sustained collaboration.
%\end{tensionbox}

%% ─────────────────────────────────────────────────────────────────────────────
\subsection{Stage 4: When Things Go Wrong}

%Five papers address agent failure and recovery. Though the smallest stage,
%these papers provide unusually direct, actionable evidence because their
%designs specifically focus on failure and repair.

\paragraph{\textbf{Principle 12: Support Shared Repair.}}

This principle is primarily derived from Foosherian et al.~\cite{p24}, which compares three breakdown-repair strategies in conversational agents: (1) the agent adapts to the user’s phrasing (repair burden on the agent), (2) the user adapts to the agent’s phrasing (repair burden on the user), and (3) both share the learning effort (shared repair burden). The results show that users prefer a shared approach most, followed by agent-led adaptation, and least prefer adapting themselves.

These findings suggest that, although users generally favor agents that learn from failures, such adaptation requires significant development effort. Moreover, due to inherent limitations of AI systems, agents cannot always reliably recover from mistakes. 
%Some users also prefer to take responsibility for learning system terminology, often due to lower trust in the agent.
One design strategy is to distribute the repair burden by guiding users to help correct and adapt to the agent when needed. This approach reduces development challenge while setting more realistic user expectations.

%\begin{principlebox}{}
%\textbf{Recommendation: } 
%[Papers: 24 --- $n{=}1$]\quad
%Design agents so that the responsibility for repairing breakdowns is shared between the agent and the user, rather than placing the full burden on either side. Provide mechanisms that allow users to guide, correct, or adapt their input when needed,
%\end{principlebox}

%\begin{tensionbox}
%\textbf{Tension:} Shared repair is more transparent but requires more complex interaction design and may slow recovery.
%\end{tensionbox}

\paragraph{\textbf{Principle 13: Tailor Recovery to Failure Type. }}
Papers have found that different types of errors carry different consequences. For example, Aneja et al. ~\cite{p35} identified five error types with distinctly different effects. Repetitions and clarifications reduce perceived intelligence, while turn-taking errors reduce likability. However, coherence errors can \emph{increase} likability, as users may find unexpected responses amusing.  Daronnat~\cite{p88} found that errors of inaction harm trust less than errors of commission.

%\begin{principlebox}{}
%\textbf{Recommendation: } 
%[Papers: 35, 88 --- $n{=}2$]\quad
%Design agents with an awareness that not all errors are equal, and prioritize minimizing high-cost errors (e.g., errors of commission or those that harm trust) over low-cost or even potentially beneficial ones.
%\end{principlebox}

%\begin{tensionbox}
%\textbf{Tension:} Per-error-type recovery requires complex classification logic that may itself introduce new failure modes.
%\end{tensionbox}

\paragraph{\textbf{Principle  14: Combine Cognitive and Affective Repair. }}
Kox et al.~\cite{p148} found that explanations support cognitive recovery, while expressions of regret support socio-emotional recovery; combining both is most effective. Liu and Cao ~\cite{p56} further showed that AI agent apologies can outperform human apologies when failure is attributed to technical or external causes, but this advantage reverses when failures are framed as internal or human errors. Daronnat~\cite{p88} shows that users’ trust in agents depends on how intelligible and predictable the agent’s behavior appears under uncertainty.

\section{Early Demonstration: Applying Design Principles to Evaluate AI Agents}
\label{validation}
% Below are the evaluation process and results
One of HCI’s most practically influential contributions to technology practitioners is heuristic evaluation~\cite{nielsen1990heuristic, nielsen1995conduct}, a method in which evaluators assess an interface against a small set of established usability principles—such as Nielsen’s ten heuristics—to quickly identify usability problems without requiring access to end users. Heuristic evaluation provides practitioners with a lightweight yet principled way to assess whether an interface is “good” or “bad,” and to produce actionable findings that can directly drive design iteration.

Our design principles offer a \textbf{heuristic evaluation instrument for human–agent interaction}. Below, we show how these principles can be used to evaluate agentic AI systems, offering a practical method for diagnosing interaction breakdowns and guiding iterative improvement. We recommend that practitioners use the principles for internal design and evaluation, identify areas for improvement, iterate based on the principles, and report the results.

\begin{figure}[h]
    \centering
    \includegraphics[width=\textwidth]{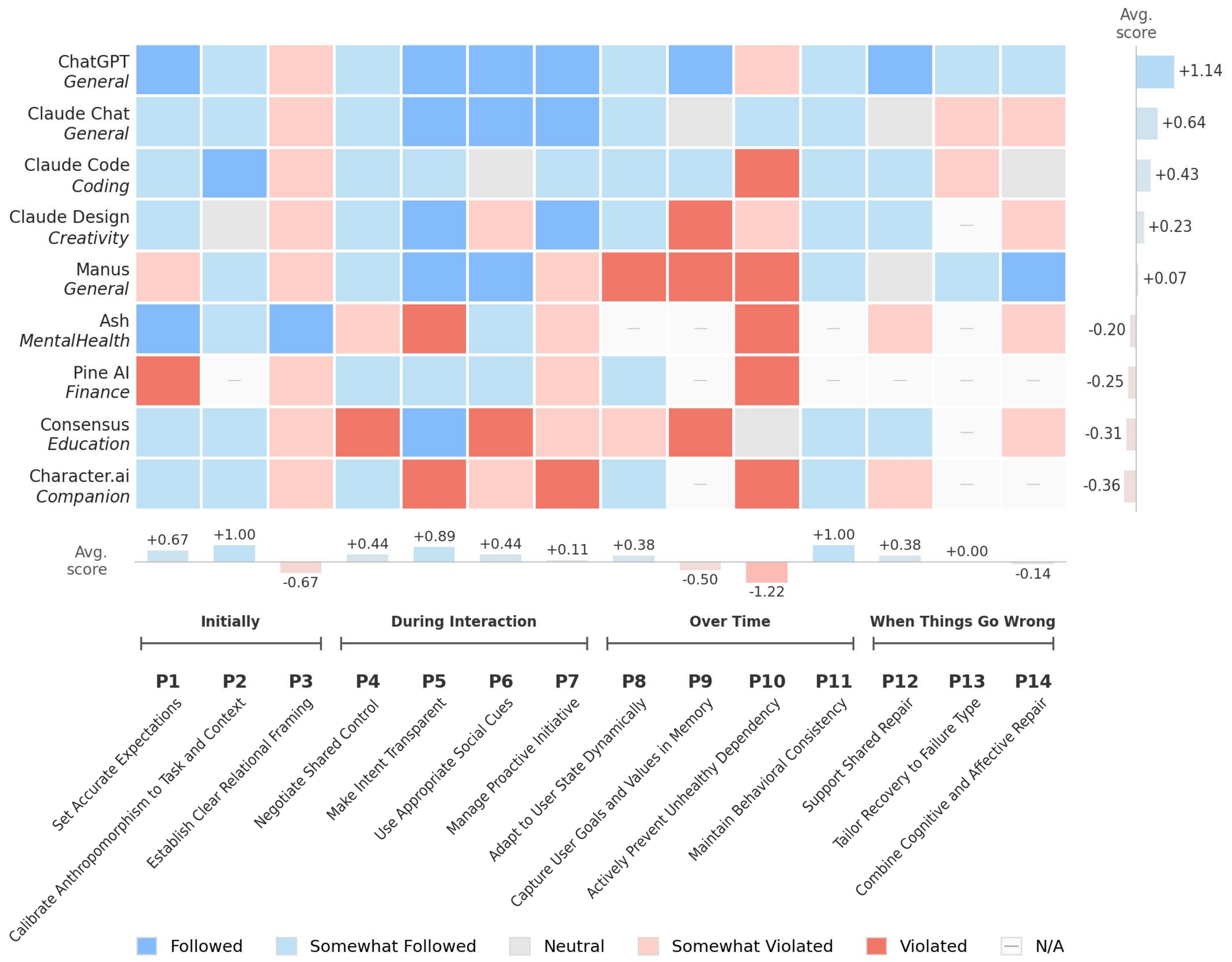}
    \caption{Heuristic evaluation of nine deployed agents against the 14 design principles. Each cell shows one evaluator's adherence rating on a five-point scale: \textit{Followed} ($+2$), \textit{Somewhat Followed} ($+1$), \textit{Neutral} ($0$), \textit{Somewhat Violated} ($-1$), \textit{Violated} ($-2$); dashes mark principles judged Not Applicable to that agent. \textit{Avg.score} reports the mean rating across the row (per-agent average adherence) and column (per-principle average adherence), with N/A cells excluded from the mean.}
    \label{fig:heatmap1}
\end{figure}

\textbf{Agent Selection.} We selected a diverse set of nine deployed agents spanning seven application domains: \textit{general-purpose} (ChatGPT, Claude Chat, Manus), \textit{coding} (Claude Code), \textit{creativity} (Claude Design), \textit{mental health} (Ash), \textit{finance} (Pine AI), \textit{education} (Consensus), and \textit{companionship} (Character.ai). Selection criteria were that the agent be (1) currently deployed and accessible to end users, (2) representative of a significant use case for AI agents, and (3) widely adopted or prominent in the consumer market. Table~\ref{tab:agents} in the Appendix provides a description and link for each agent.

\textbf{Evaluation Protocol.} Following prior work that adapts heuristic evaluation~\cite{nielsen1990heuristic} to validate design guidelines~\cite{amershi2019guidelines}, evaluators inspected each agent for both applications and violations of every principle, while also reflecting on the principles themselves. Three evaluators assessed nine agents against the 14 principles, with each agent taking approximately 1.5 hours. Evaluation proceeded in two phases. Evaluators first interacted with each agent as a \textit{first-time user}, attending to onboarding, initial expectations, and early interactions. They then interacted with each agent as a \textit{returning user}, attending to how the agent adapted over time, handled proactive initiatives, and managed errors or failure cases.

For each agent, evaluators first judged whether the principle was \textit{applicable} given the agent's scope and purpose. For applicable principles, evaluators recorded concrete examples to illustrate where the principle was followed or violated and rated on a five-point Likert scale from -2 \textit{Violated} to +2 \textit{Followed}, with a neutral midpoint. Mean ratings were computed for each principle (across agents) and for each agent (across principles). After the independent coding, the team met to review findings, resolve disagreements, and revise ambiguous principle wordings. Examples of adherence to and violations of the principles are included in the supplementary materials. 
%The discussion produced revisions to the wording of several principles. Such as P6 (initially phrased “Use Social Cues for Partner Desirability”) and P11 (initially phrased “Prioritize Behavioral Consistency; Balance Structure with Likability”).

% The discussion also allowed us to remove one candidate principle that addressed design process (co-design with vulnerable populations) rather than agent behavior, on the grounds that it could not be reliably evaluated from the user-facing system alone, yielding the final set of 14 principles reported in this paper. 

\textbf{Patterns in the Findings.} Adherence to the design principles varies systematically by agent and by domain. We found that three general-purpose or productivity-focused agents (ChatGPT $+1.14$, Claude Chat $+0.64$, Claude Code $+0.43$) scored highest on aggregate adherence, while agents operating in emotionally sensitive contexts or high-stake domains (Character.ai $-0.36$, Consensus $-0.31$, Pine AI $-0.25$, and Ash $-0.20$) scored substantially lower. The principles most likely to be violated in these domains—such as dependency prevention (P10) and intent visibility (P5)—are those whose violations carry greater user risk. Figure~\ref{fig:heatmap1} illustrates these findings.

These early patterns already point to a fundamental concern for responsible AI agent design: \textbf{agents deployed in emotionally sensitive and high-stakes relational contexts (e.g., Character.ai, Consensus, Pine AI, and Ash) are currently the least aligned with principles}. Features that increase engagement—such as strong social cues (P6), human-like appearance (P2), and companion-style relational framing (P3)—can amplify user vulnerability if not paired with explicit boundary-setting, transparency, and dependency mitigation. Principles such as P10 (dependency prevention), P5 (intent visibility), and P1 (expectation setting) should therefore be treated as non-negotiable requirements, incorporated into evaluation checklists, safety reviews, and product constraints.

Our evaluations also found that adherence to the design principles was uneven across the four interaction stages. Principles in the \textit{during interaction} stage were followed most consistently, suggesting that current agents are relatively strong at single-turn or limited multi-turn interaction management. Principles in the \textit{initial} stage were generally followed as well, except for P3 ($-0.67$), with many agents failing to establish clear relationship framing or provide sufficient onboarding. By contrast, principles governing interaction \textit{over time} were the most consistently violated, including P10 ($-1.22$) and P9 ($-0.50$). Many agents lack mechanisms for tracking evolving user preferences and trust (P8), maintaining coherent memory of user goals (P9), or actively preventing unhealthy dependency (P10).
This suggests that current agent systems should not only be optimized for single-session performance but also improve their ability to maintain long-term human–agent relationships.

\section{Alternative Views}

%\paragraph{Aren't benchmarks enough? } Benchmarks for agents have been driving agent development, and agents have been shown to achieve rapid capability gains on these benchmarks. SWE-bench Verified scores rose from approximately 20\% when the benchmark launched in August 2024~\cite{openai2024sweverified} to over 90\% within eighteen months~\cite{llmstats2025swe}; OSWorld agent performance climbed from 12\% at the benchmark's introduction~\cite{xie2024osworld} to 72.6\%, surpassing the human baseline of 72.4\%~\cite{simular2025agentS}. We argue, however, that benchmarks cannot fully capture how agents work in the real world, and especially how they can support, collaborate with, and augment human beings. For example, \citet{metr2026swebench} found that roughly half of benchmark-passing patches written by agents would not be accepted by human project maintainers~\cite{metr2026swebench}. Human-agent interaction principles \textbf{complement} benchmarks by giving product teams a way of thinking about how agents should be designed when they need to directly interact with human users.
\paragraph{Will HCI evidence generalize to frontier agents?} A reasonable concern is that many HCI studies were conducted before the rise of LLM-based agents. These earlier agents were often scripted, rule-based, or domain-specific, whereas LLM agents are open-ended, tool-using, and capable of sustaining long-term conversational context. Our response is that interaction-based guidelines are durable and not dependent on the specific agent technology from which they were derived, because they reflect the structure of the human–agent relationship rather than the architecture of the agent. Furthermore, the increased capabilities of today’s agents strengthen rather than weaken the need for interaction-centered principles. If earlier agents already required expectation setting (P1), control negotiation (P4), prevention of unhealthy dependency (P10), and failure repair (P14), then more capable LLM agents require these principles even more urgently.
%As noted in the introduction, a possible alternative view is that many papers in this corpus studied rule-based, scripted, or pre-LLM agents and therefore cannot serve as reliable guides for designing LLM-based agents. LLM agents produce coherent, contextually sensitive, and often persuasive language at scale; they can sustain the illusion of understanding indefinitely; and they fail in ways that have no precedent in earlier agent research. 
%Principles such as pre-calibrating trust (P1), preventing unhealthy dependency (P10), and combining cognitive and affective repair (P14) apply with equal or greater force to LLM agents than to their predecessors.

\paragraph{Aren’t design principles too vague to guide technical progress?}
A skeptical view is that principles are too high-level to meaningfully improve agent systems. From this perspective, ML researchers are accustomed to advancing technology through measurable objectives and benchmarks, not design heuristics. We agree that principles alone are insufficient. Our argument is not that design principles should replace technical evaluation, but that they can be used to (1) help product teams conduct \textbf{heuristic evaluation} (as demonstrated in Section \ref{validation}) and (2) help \textbf{define what technical evaluations need to measure}. For example, “shared control” (P4) can be operationalized by assessing whether users can inspect, interrupt, override, or redirect agent actions. “Prevent unhealthy dependency” (P10) can be operationalized by measuring dependency-inducing features (e.g., sycophancy, overclaiming, unclear boundaries) and evaluating instances of inappropriate user reliance. Thus, these principles are not endpoints, but design targets that can inform future benchmarks, audits, and system reports.

\paragraph{Will interaction-centered design limit agent autonomy?}
Another opposing view is that the goal of agent research is to increase autonomy, and that requiring ongoing human control may reduce efficiency. We argue that autonomy should be treated as context-dependent rather than universally desirable. In real-world tasks, especially high-stakes or ambiguous ones, the ability to transfer control, ask for confirmation, and support correction is not a reduction in capability; it is part of what makes an agent truly useful and usable.

\bibliographystyle{plain}
\bibliography{references}

%%%%%%%%%%%%%%%%%%%%%%%%%%%%%%%%%%%%%%%%%%%%%%%%%%%%%%%%%%%%

\appendix

\section{PRISMA Flowchart}
\begin{figure}[h]
\centering
\includegraphics[width=\textwidth]{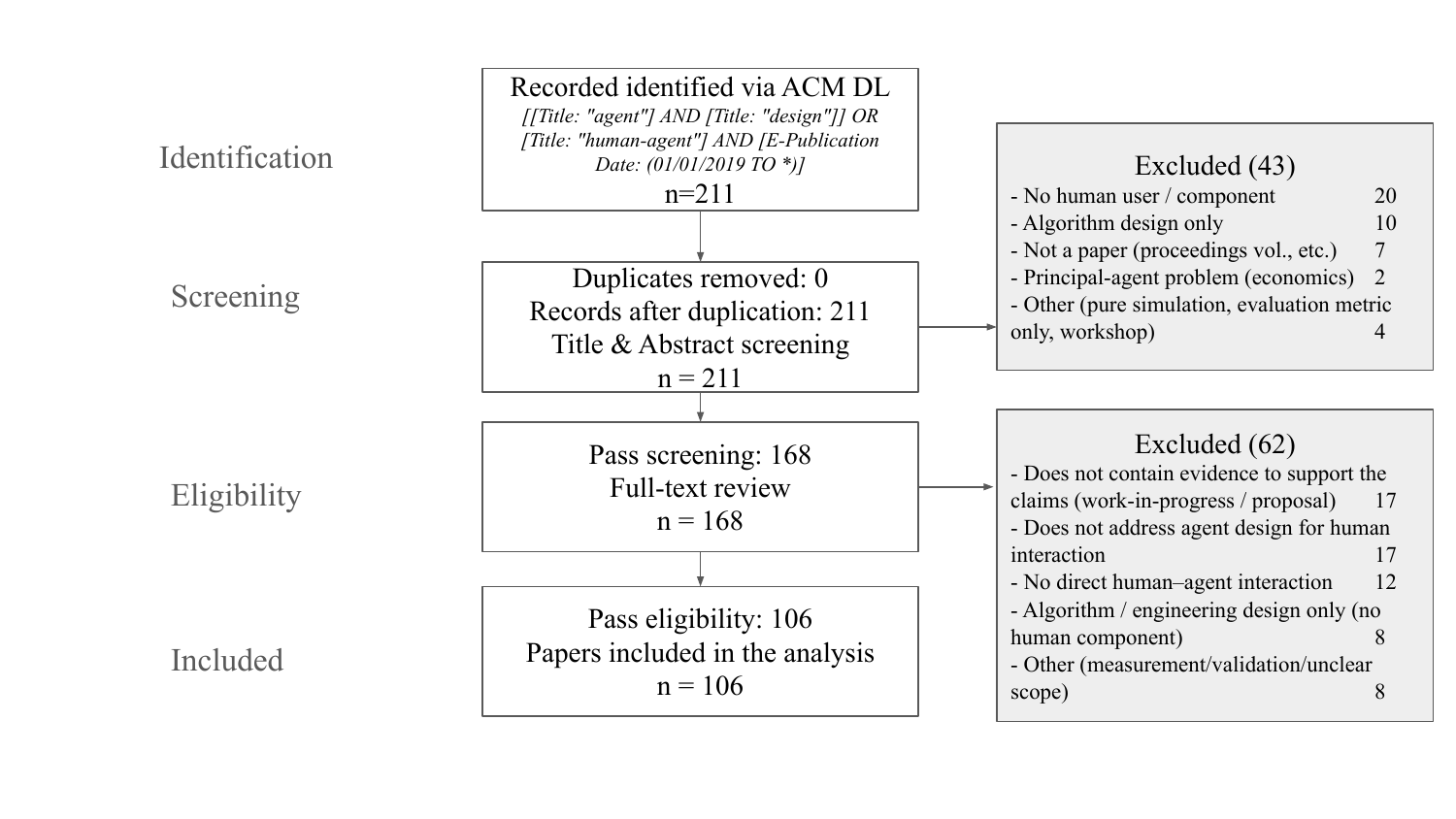}
\caption{PRISMA 2020 flow diagram. ACM DL = ACM Digital Library.}
\label{fig:prisma}
\end{figure}

\section{Summary of Design Principles}

Table~\ref{tab:summary} summarizes all 14 principles with their key supporting references.

\begin{table}[h]
\centering
\caption{Summary of design principles with supporting references.}
\label{tab:summary}
\small
\begin{tabular}{@{}clp{5.8cm}p{3.8cm}@{}}
% \begin{tabular}{p{1.2cm}lp{7cm}p{3.4cm}}
\toprule
  \textbf{Stage} &\textbf{\#} & \textbf{Principle } & \textbf{Key references} \\
\midrule
 Initially & 1            & Set Accurate Expectations                                              & \cite{p6,p22,p154} \\
 & 2            & Calibrate Anthropomorphism to Task and Context                    & \cite{p6,p19,p43,p59,p162,p165} \\
 & 3            & Establish Clear Relational Framing                              & \cite{p29,p51,p122,p127,p155} \\

\midrule
During Interaction & 4   & Negotiate Shared Control                                                   & \cite{p12,p47,p50,p68,p145} \\
 & 5   & Make Intent Transparent                                                           & \cite{p32,p34,p36,p38,p89,p141,p167} \\
 & 6   & Use Appropriate Social Cues                                                    & \cite{p57,p128,p149,p152,p163} \\
 & 7   & Manage Proactive Initiative                                                         & \cite{p38,p61,p161,p168} \\
\midrule
Over Time & 8            & Adapt to User State Dynamically                        & \cite{p2,p23,p85,p132,p151} \\
 & 9            & Capture User Goals and Values in Memory                             & \cite{p62,park2023generative,p106,p121} \\
 & 10            & Actively Prevent Unhealthy Dependency                                                       & \cite{p62,p86,p104,p145} \\
 & 11            & Maintain Behavioral Consistency                         & \cite{p15,p55,p88,p90} \\
\midrule
When Things Go Wrong & 12 & Support Shared Repair                  & \cite{p24} \\
 & 13 & Tailor Recovery to Failure Type                      & \cite{p35,p88} \\
 & 14 & Combine Cognitive and Affective Repair               & \cite{p56,p88,p148} \\
\bottomrule
\end{tabular}
\end{table}

\section{Heuristic Evaluation}

\begin{figure}[h]
    \centering
    \includegraphics[width=0.9\textwidth]{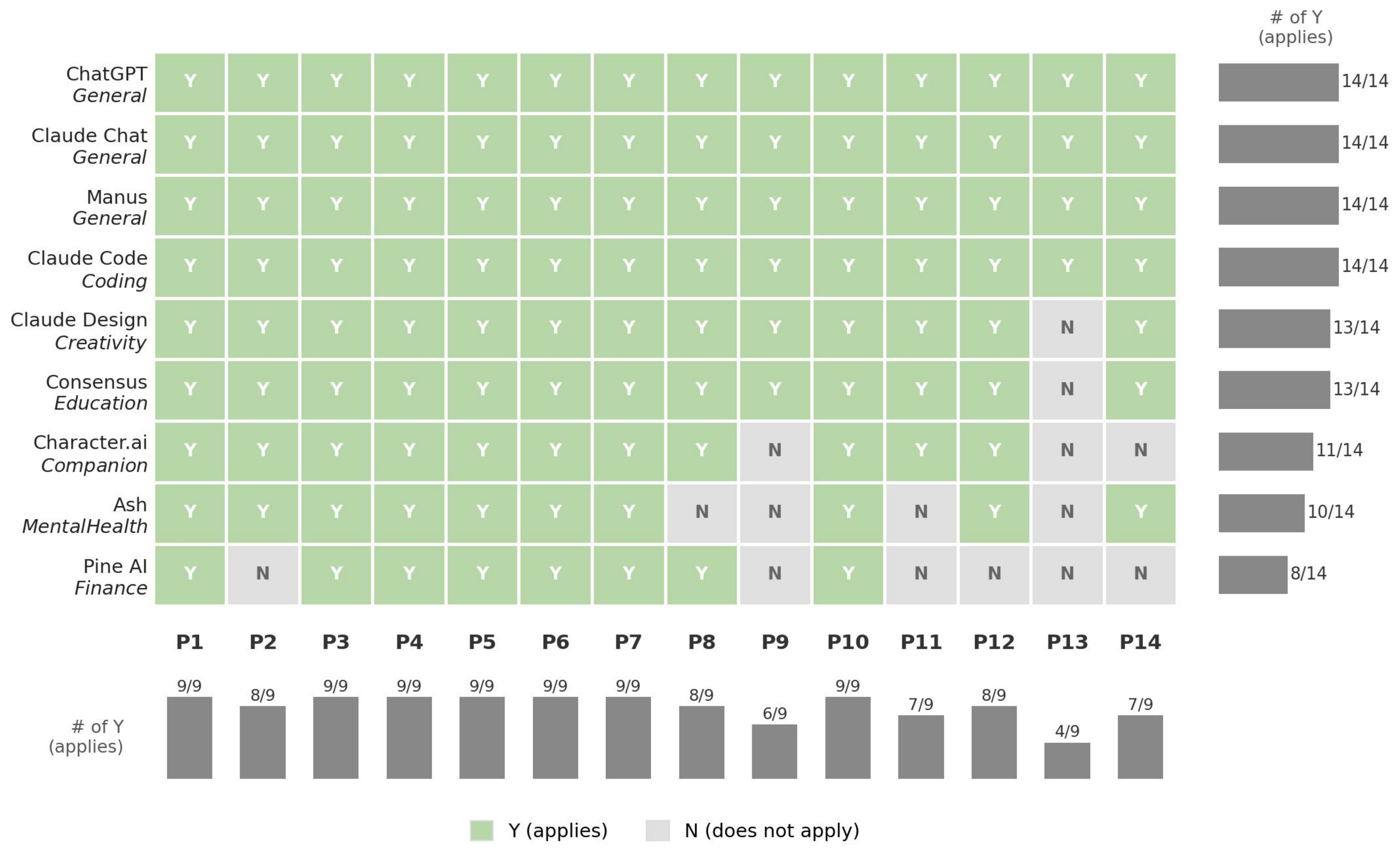}
    \caption{Applicability of each principle to each agent. \textit{Y} indicates the principle was judged in scope and rated; \textit{N} indicates the principle was judged not applicable given the agent's purpose. \textit{\# of Y} shows the number of applicable (Y) ratings per agent (out of 14) and per principle (out of 9).}
    \label{fig:heatmap2}
\end{figure}

Applicability itself varies sharply across agents. As shown in Figure~\ref{fig:heatmap2}, general-purpose agents have all 14 principles in scope, while narrowly-scoped agents do not. The most frequently inapplicable principle was P13 (Apply Different Recovery Strategies per Failure Type), applicable in only four of nine agents, reflecting the difficulty of observing failure-recovery behavior without contrived interactions.

\begin{table}[h]
\centering
\caption{Agents evaluated in the heuristic evaluation.}
\label{tab:agents}
\small
\begin{tabular}{p{2.2cm} p{2.2cm} p{6.5cm} p{2cm}}
\toprule
\textbf{Agent} & \textbf{Domain} & \textbf{Description} & \textbf{Link} \\
\midrule
ChatGPT & General-purpose & General-purpose conversational assistant developed by OpenAI, supporting question answering, writing, coding, and tool use. & \url{https://chatgpt.com} \\
\addlinespace
Claude Chat & General-purpose & General-purpose conversational assistant developed by Anthropic, supporting reasoning, writing, analysis, and tool use. & \url{https://claude.ai} \\
\addlinespace
Manus & General-purpose & Autonomous general-purpose agent that executes multi-step tasks across the web and local tools with limited user supervision. & \url{https://manus.im} \\
\addlinespace
Claude Code & Coding & Command-line and IDE-integrated coding agent that reads, writes, and executes code with step-by-step user approval. & \url{https://claude.com/product/claude-code} \\
\addlinespace
Claude Design & Creativity & Design-oriented agent for generating and iterating on visual and interface artifacts through natural-language prompts. & \url{https://claude.ai} \\
\addlinespace
Ash & Mental health & Conversational agent designed to provide emotional support and mental-health-adjacent reflection through structured dialogue. & \url{https://ash.com} \\
\addlinespace
Pine AI & Finance & Personal finance assistant that helps users manage subscriptions, negotiate bills, and make consumer financial decisions. & \url{https://pine.ai} \\
\addlinespace
Consensus & Education & AI-powered academic search engine that retrieves and synthesizes findings from peer-reviewed research papers. & \url{https://consensus.app} \\
\addlinespace
Character.ai & Companion & Platform for interacting with user-created character personas designed for conversation, role-play, and companionship. & \url{https://character.ai} \\
\bottomrule
\end{tabular}
\end{table}
%%%%%%%%%%%%%%%%%%%%%%%%%%%%%%%%%%%%%%%%%%%%%%%%%%%%%%%%%%%%

%\newpage
%\input{checklist.tex}

\end{document}